# Hydrogen and the Abundances of Elements in Impulsive Solar Energetic-Particle Events

Donald V. Reames

Institute for Physical Science and Technology, University of Maryland, College Park, MD 20742-2431 USA, email: dvreames@umd.edu

**Abstract** Hydrogen has been almost completely ignored in studies of the abundance patterns of the chemical elements in solar energetic particles (SEPs). We seek to find impulsive events where H fits these abundance patterns and document the events that do not, suggesting possible reasons for the disparity. For 24 % of the smaller impulsive SEP events, the relative abundance of H fits within one standard deviation of the power-law fit of the abundances of elements $6 \leq Z \leq 56$, relative to coronal abundances. In impulsive events with high intensities, H can be 10 to 100 times its expected value. In a few of these larger events, increased scattering at high wavenumber may preferentially detain H, perhaps with self-amplified waves; in some events pre-event proton background may contribute. In most large impulsive SEP events, however, associated shock waves must play a much greater role than previously thought; fast (>500km s$^{-1}$) coronal mass ejections contribute to 62 % of impulsive events. Shocks may sample protons from the ambient coronal plasma or residual background as well as reaccelerating heavier impulsive SEP ions injected from the region of magnetic reconnection in solar jets. Excess H may be a signature of shock acceleration.

## 1. Introduction

During decades of studies of the abundances of elements in solar energetic-particle (SEP) events, hydrogen has been ignored – the elephant in the room. Early measurements (*e.g.* Webber, 1975; Meyer, 1985) found that the average element abundances from He or C through Fe, in what we now call "gradual" SEP events, differ from solar photospheric abundances as a simple function of their first ionization potential (FIP). In addition, event-to-event variations were found to be power-law functions of the mass-to-charge ratio [$A/Q$] for elements with atomic numbers $6 \leq Z \leq 30$ in these events (Breneman and Stone, 1985). Similar power-law behavior for "impulsive" SEP events was proposed by Reames, Meyer, and von Rosenvinge (1994) based upon abundances of elements with $2 \leq Z \leq 26$. The recent extension to high $Z$ shows strong power-law behavior for $2 \leq Z \leq 82$ (Reames, 2000, 2017a; Mason *et al.*, 2004; Reames and Ng, 2004; Reames, Cliver, and Kahler, 2014a).

Why not protons? Quite simply, the dominant element H does not easily fit into the abundance patterns of the other elements, and it has therefore been ignored. Protons are studied for many aspects of SEP events (see *e.g.* Reames, 2017a), including acceleration (Lee, 1983, 2005; Temerin and Roth, 1993; Zank, Rice, and Wu, 2000; Ng and Reames, 2008; Drake *et al.*, 2009), transport (Parker, 1963; Ng, Reames, and Tylka, 1999, 2001, 2003, 2012), temporal variations (Reames, Ng, and Tylka, 2000; Reames, 2009), and energy spectra ( Reames and Ng, 2010), but not for FIP or power-law abundance patterns.

SEP events show the effects of two characteristic physical acceleration mechanisms (Reames, 1988, 1999, 2013, 2015, 2017a; Gosling, 1993). The small "impulsive" SEP events were first distinguished by huge 1000-fold enhancements of $^3$He/$^4$He (Serlemitsos and Balasubrahmanyan, 1975; Mason, 2007; Reames, 2017a). These events were soon associated with streaming non-relativistic electrons and Type III radio bursts (Reames, von Rosenvinge, and Lin, 1985; Reames and Stone, 1986); theoretically the streaming electrons could produce electromagnetic ion-cyclotron (EMIC) waves that resonate with $^3$He to produce the unique enhancements (Temerin and Roth, 1992; Roth and Temerin, 1997). Power-law enhancements of elements, also 1000-fold, between He and the heavy elements Os – Pb (Reames, 2000, 2017a; Mason *et al.*, 2004; Reames and Ng, 2004; Reames, Cliver, and Kahler, 2014a) are better explained by particle-in-cell simulations of magnetic reconnection (Drake *et al.*, 2009). Impulsive SEP events are accompanied by narrow coronal mass ejections (CMEs) that are associated with solar jets (Kahler, Reames, and Sheeley, 2001; Bučík *et al.*, 2018) involving magnetic reconnection that includes open field lines so the accelerated ions easily escape.

The largest SEP events, the "gradual" or "long-duration" events, were found by Kahler *et al.* (1984) to have a 96 % correlation with fast, wide CMEs that drive shock waves out from the Sun. These shock waves accelerate the SEPs (Lee, 1983, 2005; Zank, Rice, and Wu, 2000; Cliver, Kahler, and Reames, 2004; Ng and Reames, 2008; Rouillard *et al.*, 2012; Lee, Mewaldt, and Giacalone, 2012; Desai and Giacalone, 2016; Reames, 2017a), usually requiring shock speeds above 500–700 km s$^{-1}$ (Reames, Kahler, and Ng, 1997), and the accelerated ions are widely distributed in space (Reames, Barbier, and Ng, 1996; Cohen, Mason, Mewaldt, 2017; Reames, 2017b). Generally, the shock waves sample the 1–3 MK coronal plasma (Reames, 2016a, 2016b, 2018b), but often they reaccelerate suprathermal ions left over from the ubiquitous small impulsive SEP events (Mason, Mazur, and Dwyer, 2002) which serve as a pre-accelerated seed population injected into the shock (Desai *et al.*, 2003; Tylka *et al.*, 2005; Tylka and Lee, 2006; Bučík *et al.*, 2014, 2015, 2018; Chen *et al.*, 2015). The reacceleration of impulsive suprathermal ions makes it somewhat difficult to distinguish impulsive and gradual events from element abundances alone, but the shock waves seem to diminish the enhancement of abundance ratios such as Fe/O so they are usually distinguishable from values in pure impulsive events (Reames, 2016b, 2019b).

When we average element abundances over many gradual SEP events and divide by the corresponding abundances of the solar photosphere, we find (*e.g.* Meyer, 1985; Reames, 1995, 2014, 2017a) that the low-FIP (< 10 eV) elements, which are ions in the photosphere, are enhanced by a factor of about four relative to high-FIP elements, which are neutral atoms there. The ions are apparently aided in their journey up into the corona by Alfvén waves (Laming, 2009, 2015), while the neutral atoms are not. The "FIP effect" seen in SEPs differs somewhat from that of the solar wind (Mewaldt *et al.*, 2002; Desai *et al.*, 2003; Schmelz *et al.*, 2012; Reames, 2018a) perhaps because of differing open- and closed-field origins of the two coronal samples (Reames, 2018a, 2018b). The reference coronal SEP abundances are listed in Appendix A.

We use the term "enhancement" for the ratio of the observed to the average coronal-reference abundances. These abundance enhancements have been found to vary as a power-law function of *A/Q* (Breneman and Stone, 1985; Reames, 2016a, 2016b, 2018b) from event to event and as a function of time during events, largely because of particle transport from the source. Power-law dependence of ion scattering upon magnetic rigidity during acceleration or transport (Parker, 1963; Ng, Reames, and Tylka, 1999, 2001, 2003, 2012; Reames, 2016a) results in a power-law dependence upon *A/Q* when ion abundances are compared at the same particle velocity. If Fe scatters less than O in transit, Fe/O will increase early in an event and be depleted later on; this dependence on time and space, along the field ***B***, becomes longitude dependence because of solar rotation. In the case of small impulsive events, the strong *A/Q* dependence probably comes directly from acceleration in islands of magnetic reconnection in the source (*e.g.* Drake *et al.*, 2009).

In early studies of element abundances in impulsive SEP events, Reames, Meyer, and von Rosenvinge (1994) saw groupings of abundances, with He, C, N, and O being unenhanced relative to the reference coronal abundances, the elements Ne, Mg, and Si were all enhanced by a factor of ≈ 2.5, and Fe was enhanced by a factor of ≈ 7. In their interpretation, C, N, and O were almost fully ionized, like He, with *A/Q* ≈ 2, while Ne, Mg, and Si were in a stable state containing two orbital electrons. This pattern occurs at an electron temperature of *T* = 3–5 MK. At a lower *T*, O would begin to pick up electrons and at a higher *T*, Ne would become fully ionized like He. More recently, Reames, Cliver, and Kahler (2014a) compiled and studied a list on 111 impulsive SEP events occurring over nearly 20 years, identifying their solar source events, and Reames, Cliver, and Kahler (2014b) developed a best-fit scheme to determine the source temperature and the power-law fit of abundance enhancements *versus A/Q* for each event. Each *T* implies a *Q*-value and an *A/Q* for each element and the fit of enhancement *versus A/Q* yields a $\chi^2$-value for that *T*. For each impulsive SEP event we choose the fit, and the associated *T*, with the minimum $\chi^2$. Temperatures for nearly all of the events were found to fall in the 2.5–3.2 MK range (Reames, Cliver, and Kahler, 2014b, 2015), so the differences in *T* between impulsive events is not very significant. For gradual events, however, the range of *T* is 0.8 to 3.2 MK, and is helpful in distinguishing events with large components of impulsive suprathermal ions. The recent article by Reames (2019a) shows a detailed example of the application of this technique to three impulsive SEP events and discusses the presence of He in the abundance scheme of impulsive SEP events. The source values of He/O for gradual SEP events are found to vary rather widely from 30 to 120 (Reames, 2017c, 2018b); perhaps these variations result from delay and difficulty in ionizing and transporting He to the corona, because of its uniquely high FIP (Laming, 2009). Impulsive SEP events mirror these variations, but they also show a unique sample of "He-poor" events of unknown origin with the observed value of He/O as low as $\approx 2$.

What about H? To what extent can we organize the abundance of H within the scheme of the other elements? We begin with the 111 impulsive events in the list of Reames, Cliver, and Kahler (2014a). However, only 70 of these events have proton intensities above background in the 2–3MeV region where we have measurements. We use observations made by the *Low-Energy Matrix Telescope* (LEMT) onboard the *Wind* spacecraft, near Earth (von Rosenvinge *et al.*, 1995; Reames *et al.*, 1997; see also Chapt. 7 of Reames, 2017a). LEMT primarily measures elements He through Pb in the region of 2–20MeV amu$^{-1}$. The elemental resolution of LEMT up through Fe is shown in detail by Reames (2014). LEMT resolves element groups above Fe as shown by Reames (2000, 2017a). Unlike other elements, the protons on LEMT are limited to a single sampled interval bounded by the front-detector (dome) threshold (von Rosenvinge *et al.*, 1995; Reames *et al.*, 1997).

Throughout this text, whenever we refer to the element He or its abundance, we mean $^4$He, unless $^3$He is explicitly stated.

## 2. Small Impulsive SEP Events

We begin with some of the smallest of our impulsive SEP events. The particles in these small events are generally regarded to travel scatter-free (*i.e.* scattering mean free path [$\lambda$] > 1 AU) based upon careful studies of these events by Mason *et al.* (1989). Under these circumstances the observed element abundances should directly sample the source.

Figure 1 shows principal ion intensities and abundance ratios as a function of time and the ion enhancements *versus A/Q* for two small impulsive SEP events. The least-squares fits only include the enhancements of elements with $Z \geq 6$; H and He are not included in the fits, but this fitted line is extrapolated to $A/Q = 1$ and the error of extrapolation is plotted. The fit to H and He is coincidental.

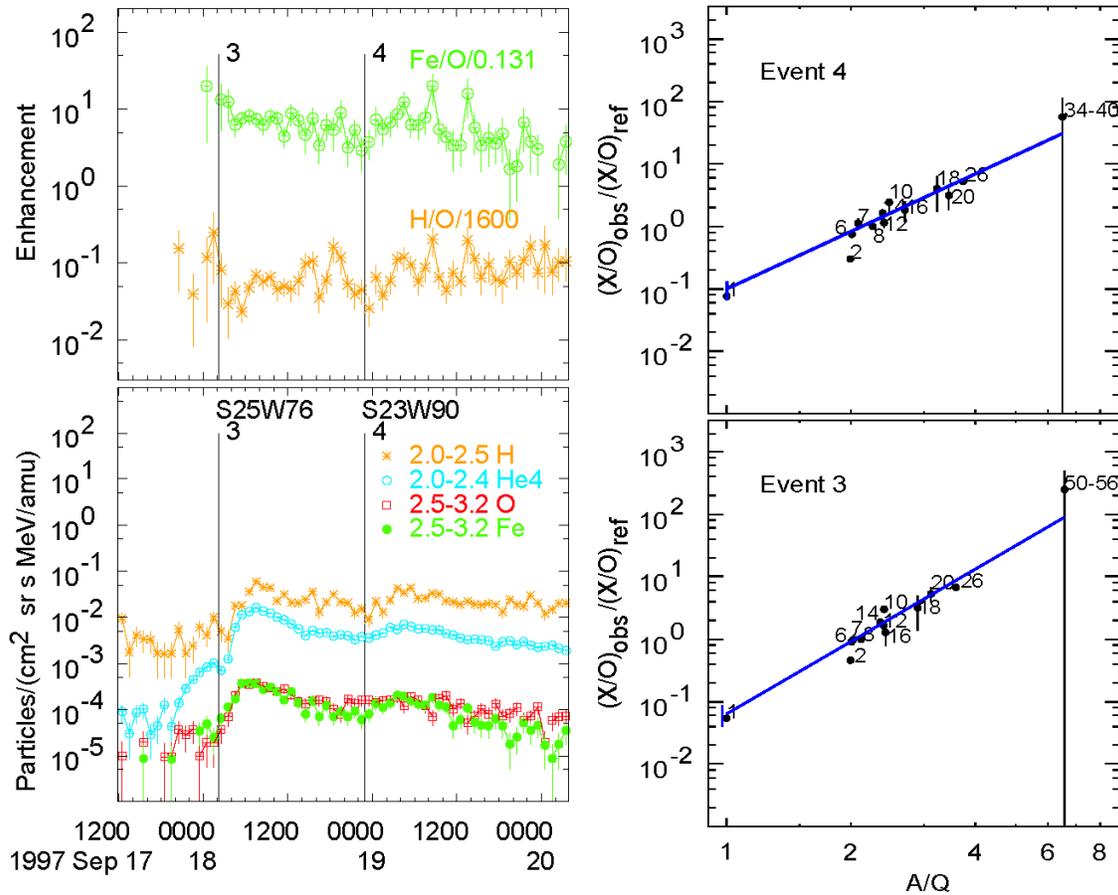

**Figure 1**. Intensities of H, He, O, and Fe (**lower-left**) and normalized abundance enhancements H/O and Fe/O (**upper-left**) are shown *versus* time for two impulsive SEP events. Event numbers 3 and 4, marking the event onset times, refer to the list of Reames, Cliver, and Kahler (2014a); the solar source locations are also listed. The **right panels** show enhancements, labeled by *Z*, *versus A/Q* for each event, with best-fit power law for elements with $Z \geq 6$ (*blue line*) extrapolated down to $A/Q = 1$.

Intensities, enhancements, and power-law fits are shown for two more small events from solar cycle 24 in Figure 2. Event 105, shown in the left panels of Figure 2, is a slowly evolving event while event 107, in the right panels, has a fast rise and rapid decay. Again, the power-law fit to H and He is coincidental. What all four events (in Figures 1 and 2) share is low proton intensities of <0.1 (cm$^2$ sr s MeV)$^{-1}$ at ≈2 MeV.

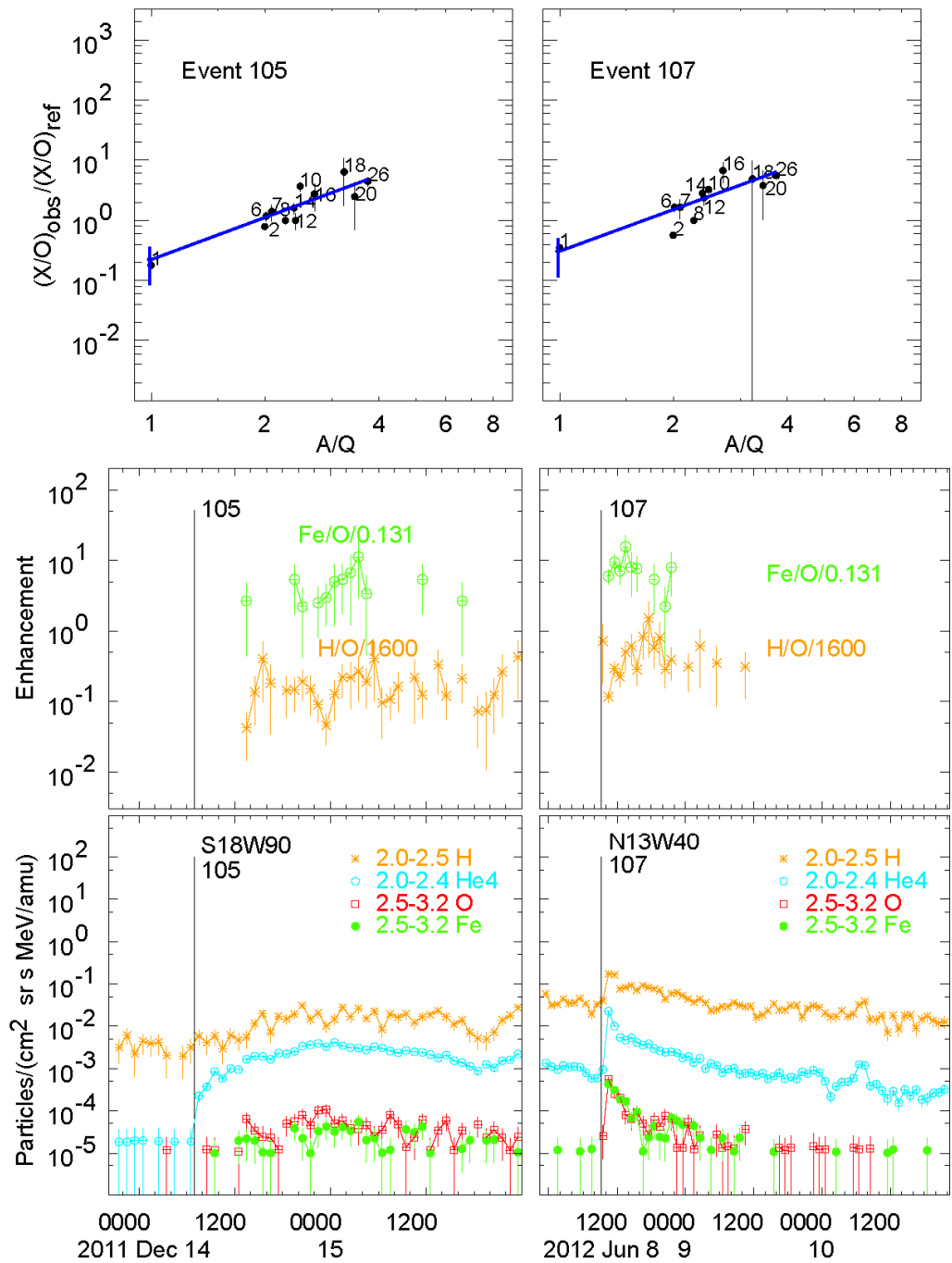

**Figure 2**.  Intensities of H, He, O, and Fe (**lower-panels**) and normalized abundance enhancements H/O and Fe/O (**center-panels**) are shown *versus* time for two impulsive SEP events.  Event numbers 105 and 107, marking the event onset times, refer to the list of Reames, Cliver, and Kahler (2014a); the solar source locations are also listed. The **upper panels** show enhancements, labeled by *Z*, *versus A/Q* for each event, with best-fit power law for elements with $Z \geq 6$ (*blue line*) extrapolated down to $A/Q = 1$.

## 3. He-Poor Events

He-poor impulsive SEP events were originally noted by Reames, Cliver, and Kahler (2014a) and have been discussed as a function of energy in the context of power-law fits by Reames (2019a). In our current context, the events are small, quite apart from their unusual suppression of He. Figure 3 shows a pair of He-poor events where the power-law fits in the right panels are extrapolated below He to $A/Q = 1$. While H exceeds the extrapolated value in event 34, perhaps affected by H background from an event ≈ four hours earlier, Event 35 agrees well with the expected value, despite the suppressed He, and other He-poor events (*e.g.* Event 79) behave like Event 35.

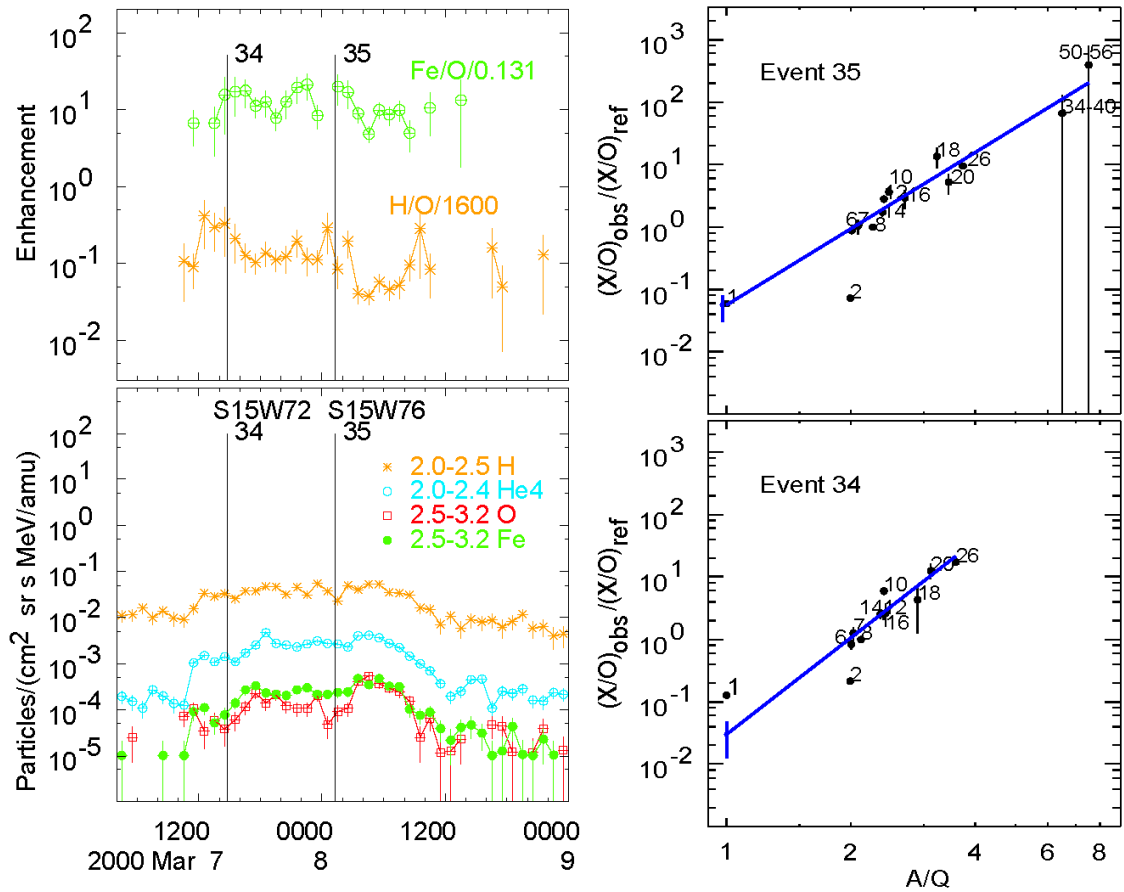

**Figure 3**. Intensities of H, He, O, and Fe (**lower-left**) and normalized abundance enhancements H/O and Fe/O (**upper-left**) are shown *versus* time for two He-poor impulsive SEP events. Event numbers 34 and 35, marking the event onset times, refer to the list of Reames, Cliver, and Kahler (2014a); the solar source locations are also listed. The **right panels** show enhancements, labeled by $Z$, *versus* $A/Q$ for each event, with best-fit power law for elements with $Z \geq 6$ (*blue line*) extrapolated down to $A/Q = 1$.

## 4. Large Impulsive SEP Events

There are two classic large impulsive SEP events that have been discussed in the literature: Event 37 on 1 May 2000 (*e.g.* Reames, Ng, and Berdichevsky, 2001) and Event 49 on 14 April 2001 (*e.g.* Tylka *et al.*, 2002). These events are shown in Figure 4.

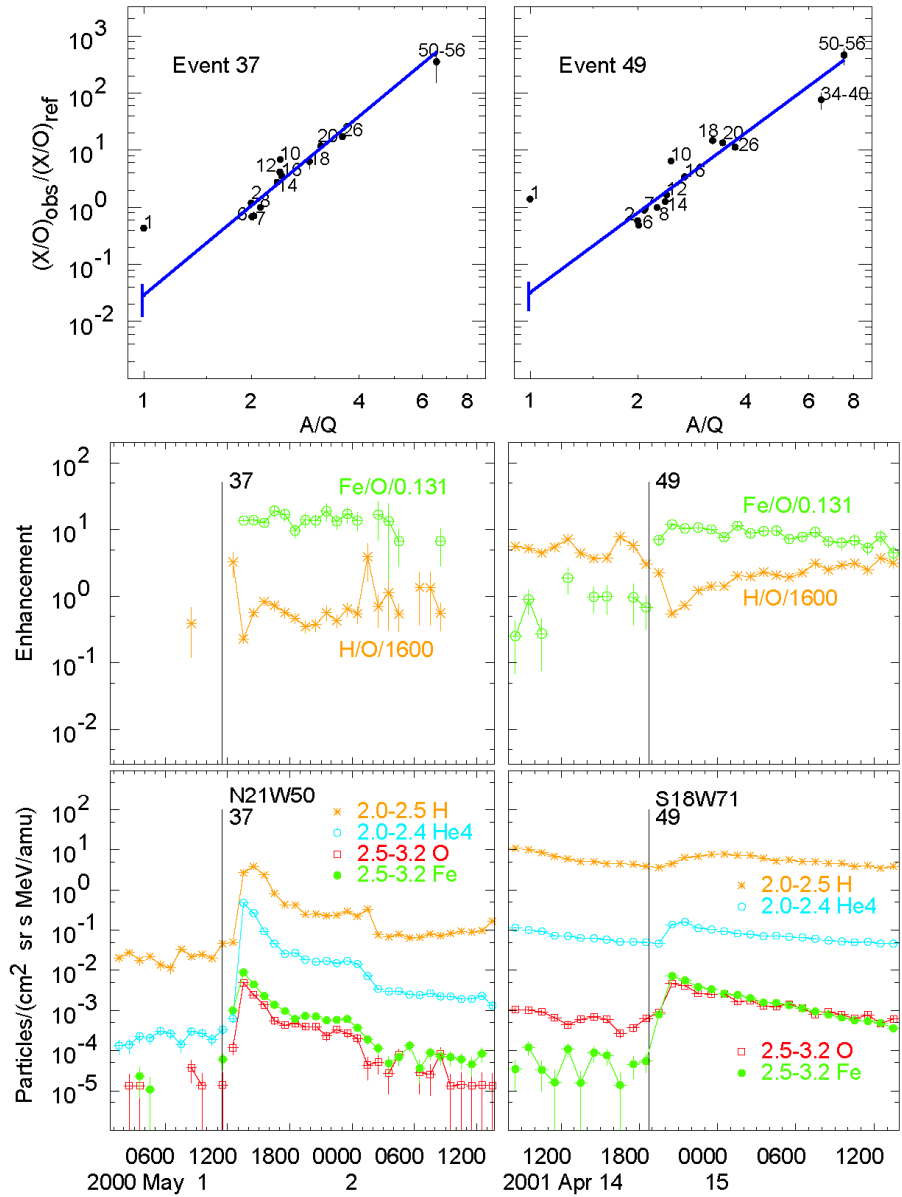

**Figure 4**. Intensities of H, He, O, and Fe (*lower-panels*) and normalized abundance enhancements H/O and Fe/O (**center-panels**) are shown *versus* time for two large impulsive SEP events. Event numbers 37 and 49, marking the event onset times, refer to the list of Reames, Cliver, and Kahler (2014a); the solar source locations are also listed. The **upper panels** show enhancements, labeled by $Z$, *versus* $A/Q$ for each event, with best-fit power law for elements with $Z \geq 6$ (*blue line*) extrapolated down to $A/Q = 1$.

For the events in Figure 4, the proton enhancement exceeds the expected value by an order of magnitude. For Event 37, pre-event proton background certainly seems an unlikely issue, since the proton intensity increases from 0.049±0.007 (cm$^2$ sr s MeV)$^{-1}$ before the event to 3.9±0.07 (cm$^2$ sr s MeV)$^{-1}$ at the peak. However, higher proton intensities can amplify Alfvén waves which begin to scatter the particles (Ng, Reames, and Tylka, 1999, 2003). Particles of rigidity $P$ and pitch angle $\mu$ resonate with waves of wave number $k = B/\mu P$ in a magnetic field strength $B$. Streaming protons of 2 MeV amplify their own resonant waves, but waves that resonate with He of 2 MeV amu$^{-1}$ are produced by 8 MeV protons, which are much less numerous. Angular distributions of various ions during impulsive Event 37 are shown in Figure 5.

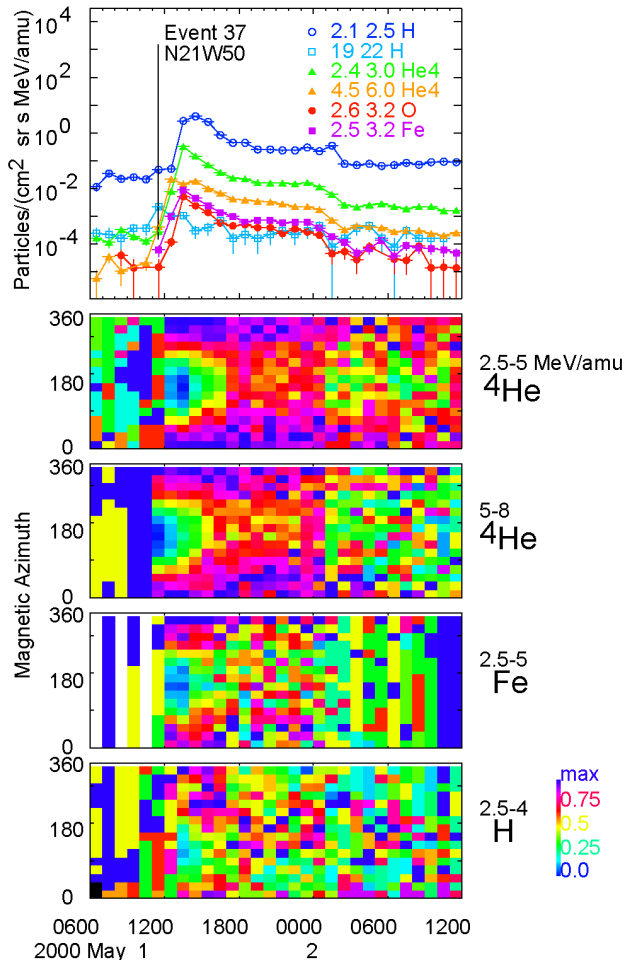

**Figure 5**. Angular distributions of H, He (in two energy intervals) and Fe are shown *versus* time for Event 37 in the **lower panels** while several ion intensities are shown in the **upper panel**. Particle arrival directions are binned relative the direction of ***B*** which is outward from the Sun during this event.

Particle arrival directions in LEMT are binned relative to the direction of ***B*** so that outward-flowing ions such as He and even Fe peaks at 0° early in this event and continue outward flow throughout the event. Proton flow is less certain early in the event and even shows sunward flow after 1800 UT on 1 May. Proton scattering is also shown by the broadened peaks in the lower panels of Figure 4, for ≈ 2 MeV protons but not for ≈ 2 MeV amu$^{-1}$ ions. If the protons are detained they are sampled more often.

However, scattering cannot explain the H enhancement in all of the large impulsive events, such as Event 54 shown in Figure 6.0

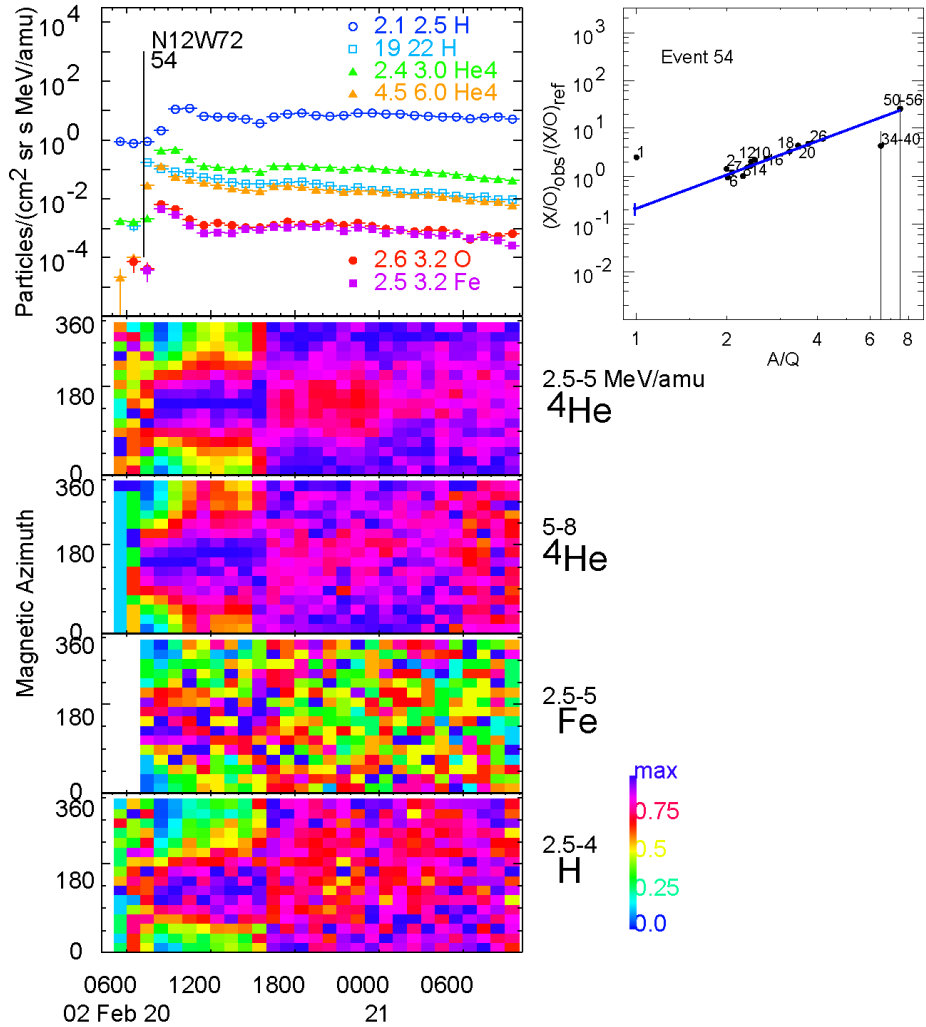

**Figure 6**. Angular distributions of H, He (in two energy intervals) and Fe are shown *versus* time for Event 54 in the **lower panels** while several ion intensities are shown in the **upper-left panel**. The **upper-right panel** shows enhancements, labeled by Z, *versus* A/Q for this event, with best-fit power law for elements with Z ≥ 6 (*blue line*) extrapolated down to A/Q = 1. Particle-arrival directions are binned relative the direction of ***B*** which is sunward early in this event so the outward flowing ions peak at 180°; the direction of ***B*** reverses at about 1700 UT on 20 February while the ions continue their outward flow.

For Event 54 in Figure 6, protons stream out from the Sun just as well as the other ions. When the field polarity reverses at about 1700 UT on 20 February, the ion anisotropies are reduced but their outward flow continues. The proton intensity peak is also spread in time somewhat in this event, but the protons flow away as easily as the ions.

## 5. Distributions and Properties of Impulsive SEP Events

We have already indicated that size, *i.e.* proton or He intensity, is a factor in the relative enhancement of H in these events and that proton scattering may sometimes be a consideration. To study effects of scattering, Figure 7a shows the distribution of the 70 events with measurable protons with the abcissa as the ratio of the observed H enhancement divided by the expected value extrapolated from the fit of the ions with $Z \geq 6$ and the ordinate as front-to-back azimuth of the proton flow during the first six hours of the event. The symbol size and color of each event depends upon peak proton intensity. An observed/expected ratio =1 means that the H enhancement lies on the fit line of the other elements; this value is shown as a solid line in Figure 7. Figure 7b shows how the observed/expected H ratio varies with event size, *i.e.* with peak proton intensity, and Figure 7c shows the overall histogram of the distribution.

In Figure 7, 17 events (24 %) of the impulsive SEP events have proton intensities within one standard deviation of the value predicted from the least-squares fit of the elements with $Z \geq 6$ and 44 events (63 %) are within two standard deviations. However, it is clear that the small events fit rather well while the larger, more intense events, beginning with peak proton intensities $> 0.1$ (cm$^2$ sr s MeV)$^{-1}$, contribute to a broad maximum centered about ten times above the expected H intensity.

Recent studies of H enhancements in gradual SEP events (Reames, 2019b) show clearly that those gradual SEP events with positive powers of *A/Q*, like impulsive SEP events, tend to have H abundances a factor of ten or more above expectation, while most gradual SEP events, with negative powers of *A/Q*, tend to have H enhancements nearer expectation. For gradual SEP events, these two distributions can also be distinguished based upon source plasma temperature. Our impulsive SEP events here all have similar positive powers of *A/Q* and source temperatures, so they are less easy to distinguish.

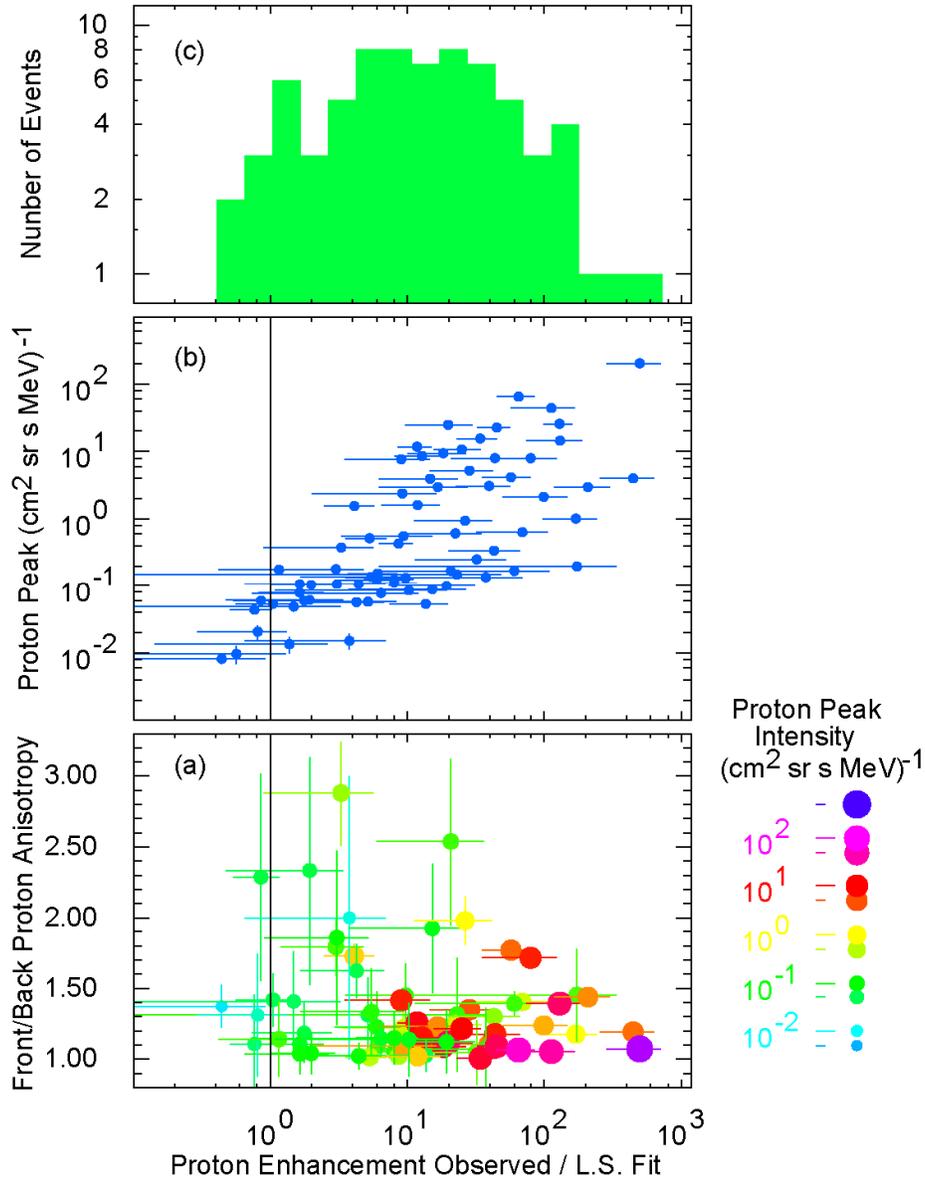

**Figure 7**. **Panel a:** the front/back directional anisotropy of protons during the first six hours is shown *versus* the observed enhancement of H relative to that expected from the power-law fit for elements $Z \geq 6$ for 70 impulsive SEP events; the symbol size and color show the peak proton intensity of each event. **Panel b:** the peak proton intensity *versus* the observed/expected H ratio for each event, and **panel c** shows a histogram of the distribution of observed/expected H enhancement ratio. An observed/expected H enhancement =1 is shown as a *solid line* in **panels a and b.**

For the larger events, shock acceleration could be a consideration, so we replot the lower panels of Figure 7 using CME speed from the table in Reames, Cliver, and Kahler (2014a) as the symbol for each event in Figure 8.

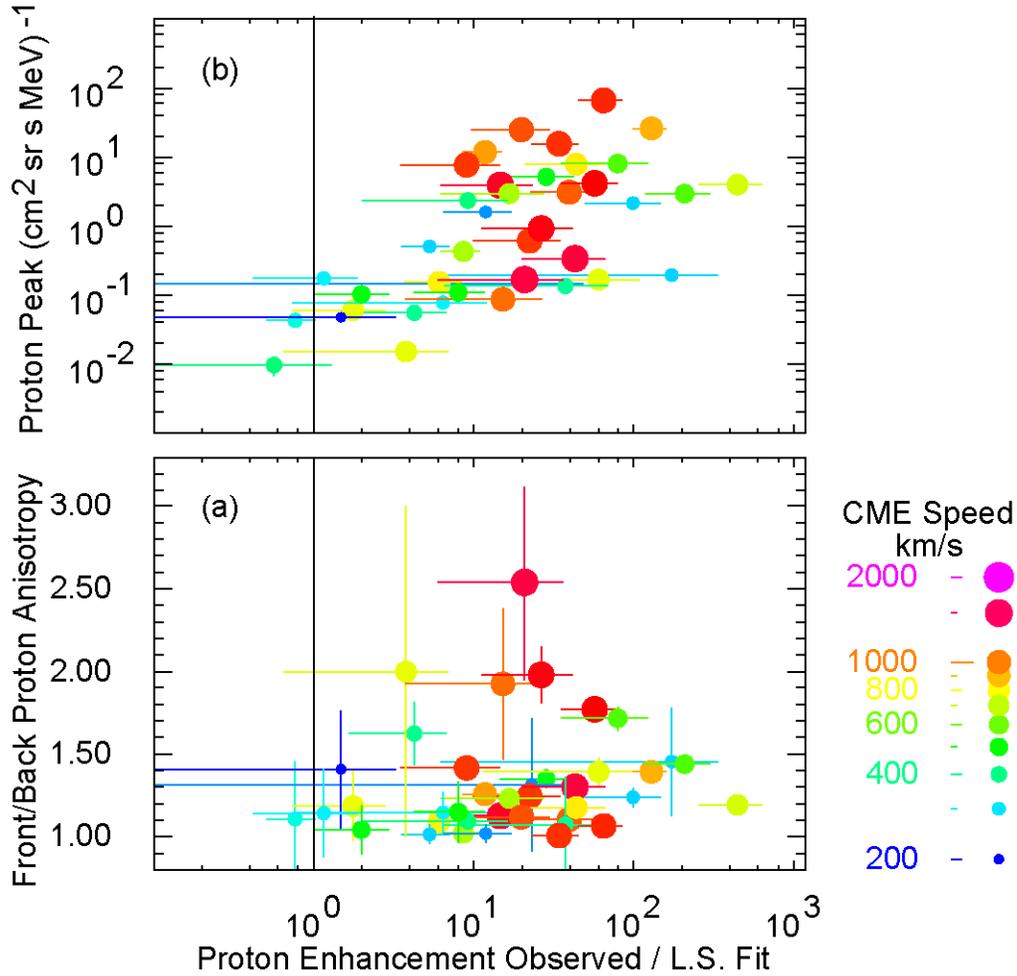

**Figure 8**. **Panel a:** the front/back directional anisotropy of protons during the first six hours is shown *versus* the enhancement of H relative to that expected from the power-law fit for elements $Z \geq 6$ for 39 impulsive SEP events with CMEs. **Panel b:** the peak proton intensity *versus* the observed/expected H for each event. In *both panels* the *symbol size* and *color* depend upon the associated CME speed. An observed/expected H enhancement =1 is shown as a *solid line* in *both panels*.

Unfortunately we now have only 40 events with CME measurements in our sample, although most of the unassociated events are small. Reames, Kahler, and Ng (1997) found that essentially all CMEs with speeds above 750km s$^{-1}$ accelerated SEPs and most of those with $500 \leq V_{CME} \leq 750$km s$^{-1}$ did. Here, 18 of the 39 events (45 %) have $V_{CME} \geq$ 750km s$^{-1}$ and 25 (62 %) have $V_{CME} \geq 500$km s$^{-1}$. However, of the 15 (37 %) events have $V_{CME} < 500$ km s$^{-1}$, only two of them (Events 52 and 55) are more than two standard deviations above the proton enhancement predicted from the least-squares fit from the ele-ments with $Z \geq 6$; both of these events have high or even rising pre-event proton background. Generally the large events with fast CMEs have the ten-fold excess of protons.

## 6. Discussion

For small impulsive SEP events, H fits on the increasing power-law trend line of the abundance enhancements of the other elements. H fits even when He does not, as in the He-poor events. He is bypassed. This suggests that the He suppression in He-poor events exists in the source plasma and is unrelated to the physics of acceleration. Perhaps He is poorly conducted into the corona because of its high FIP (Laming 2009); however, Ne and Ar seem quite plentiful.

While disproportionate scattering may detain H in some large SEP events and self-amplification of waves may contribute, scattering also depends upon preexisting conditions and this is not likely to be a significant contributor to the excess enhancement of H. An analysis of scattering is not particularly helpful in explaining H abundances. However, the early analysis of scatter-free transport in impulsive SEP events (Mason *et al.*, 1989) was made using He, not H.

The most probable contributor to excess enhancement of H in large impulsive SEP events is acceleration by the CME-driven shock wave of the direct ambient coronal plasma or pre-existing background SEPs as well as pre-accelerated impulsive ions from magnetic reconnection in the jet. Figure 9 suggests a way that these components might contribute to Event 54 (shown previously in Figure 6). The power law describing the impulsive component (blue) comes from the fit of the enhancements observed elements with $Z \geq 6$. Protons from this component contribute almost nothing. The power law describing the ambient component (red) includes H and assumes that both components contribute similar amounts of He. The ambient abundances of $Z \geq 6$ ions (only uncertain C and O are shown) fall on this line at higher values of *A/Q* since they probably have lower *T*, perhaps $\approx$ 1MK, instead of 2.5MK for the impulsive ions. The shock wave in Event 54 has a speed of 954 km s$^{-1}$(Reames, Cliver, and Kahler, 2014a). The decreasing slope of the red line in Figure 9 has probably also caused a decrease in the slope of the impulsive component relative to its initial value at the jet source (Reames, 2016b, 2018b).

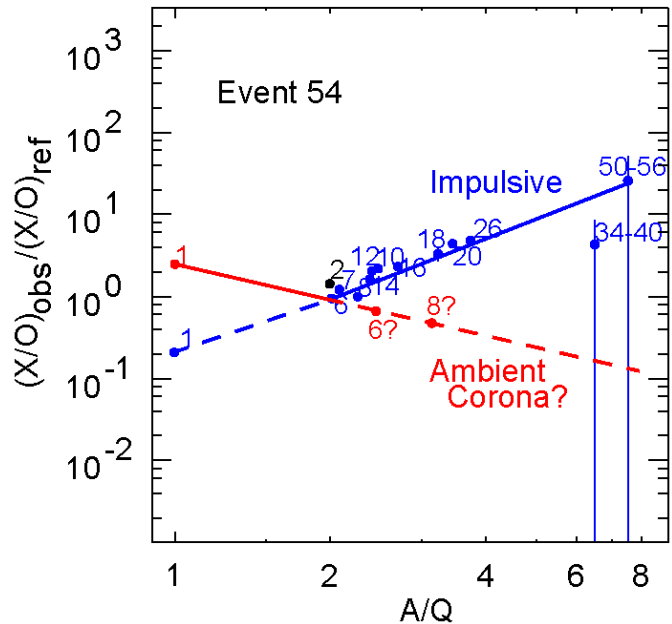

**Figure 9.** Element enhancements, labeled by *Z*, *versus A/Q* for Event 54 (see Figure 6), together with possible sources for shock acceleration from pre-accelerated impulsive ions (*blue*) and ambient corona or pre-event plasma (*red*).  Helium is assumed to receive comparable contributions from both sources in this event.

    We have assumed that the red line in Figure 9 passes near He and that the ambient component contributes to it, as may be the case in several events with elevated He.  If not, the descent of the red line is undefined, and it may be much steeper.  However, the line cannot pass above He.  He is an unfortunate fulcrum for balancing the two possible components of the seed population, since its abundance can be quite variable (Reames, 2019a).  There may also be events where the ambient corona dominates the C and O observations as well, but such events would have reduced values of Fe/O and would not be included with "impulsive" events.  In any case, shocks treat the two components of the seed population independently; both are present in the output, at some level, if they are present in the input.

    We should note that a possible component of the seed population for shock acceleration includes residual ions from previous SEP events.  The presence of a high proton background prior to an event may contribute a source which can replace "ambient corona" in Figure 9, or simply add a third component, increasing the complexity of the observations.  However, the observed proton excess does not scale with the proton background.  Pre-event proton background is higher for small SEP events, yet it is the large impulsive events, with minimal background, that have the ten-fold excesses of protons.

If this analysis is correct, the presence of an excessive proton enhancement, whatever the proton source, may, in fact, be evidence of the presence of shock acceleration in events that are otherwise dominated by characteristics of impulsive SEP events. Shock acceleration requires streaming protons to generate the waves that scatter resonant ions back and forth across the shock (Lee, 1983, 2005; Ng and Reames, 2008). Shocks cannot accelerate impulsive suprathermal ions without an adequate source of protons, which always dominate the absolute intensities and the wave amplification.

Clearly, the possibility of multiple components to the seed population also applies to shock acceleration in gradual SEP events as well. Thus it is not surprising that there are events where a single power law does not fit at any source temperature. The combination of multiple components and variations with time complicates some gradual events but also explains the origin of their complexity.

In summary:

i) In small impulsive SEP events, H fits the power-law dependence of the heavier ions.

ii) The small He-poor events are no exception. In these events, only the He abundance is unique and it must be a property of the source, not the acceleration.

iii) Larger impulsive SEP events show excess H abundance, probably accelerated by a CME-driven shock wave from ambient coronal material or other residual ions along with the "impulsive" pre-accelerated ions with $Z \geq 2$ from magnetic reconnection in the associated solar jet. Shocks sample all components of the seed population independently but protons are required to produce waves.

iv) Multiple components in the seed population could lead to regions of *A/Q* with different power-law enhancement behavior. A single power law may not always fit.

**Acknowledgments:** The author thanks Steve Kahler for helpful discussions on the subject of this manuscript.

## Disclosure of Potential Conflicts of Interest

The author declares he has no conflicts of interest.

## Appendix A: Reference Abundances of Elements

The average element abundances in gradual SEP events are a measure of the coronal abundances sampled by SEP events (Reference gradual SEPs in Table 1). They differ from photospheric abundances (Table 1) by a factor that depends upon FIP (*e.g.* Reames 2018a; 2018b). Ion "enhancements" are defined as the observed abundance of a species, relative to O, divided by the reference abundance of that species, relative to O.

**Table 1** Reference element abundances

|       | Z     | FIP [eV] | Photosphere[1]         | Reference Gradual SEPs[2] | Avg. Impulsive SEPs[3] |
|-------|-------|----------|------------------------|---------------------------|-------------------------|
| H     | 1     | 13.6     | $(1.74\pm0.17)\times10^{6}$ [*] | $(1.6\pm0.2)\times10^{6}$ | -                       |
| He    | 2     | 24.6     | 156000±7000            | 57000±5000                | 53000±3000[**]          |
| C     | 6     | 11.3     | 550±76[*]              | 420±10                    | 386±8                   |
| N     | 7     | 14.5     | 126±35[*]              | 128±8                     | 139±4                   |
| O     | 8     | 13.6     | 1000±161[*]            | 1000±10                   | 1000±10                 |
| Ne    | 10    | 21.6     | 210±54                 | 157±10                    | 478±24                  |
| Mg    | 12    | 7.6      | 64.5±9.5               | 178±4                     | 404±30                  |
| Si    | 14    | 8.2      | 61.6±9.1               | 151±4                     | 325±12                  |
| S     | 16    | 10.4     | 25.1±2.9[*]            | 25±2                      | 84±4                    |
| Ar    | 18    | 15.8     | 5.9±1.5                | 4.3±0.4                   | 34±2                    |
| Ca    | 20    | 6.1      | 4.0±0.7                | 11±1                      | 85±4                    |
| Fe    | 26    | 7.9      | 57.6±8.0[*]            | 131±6                     | 1170±48                 |
| Zn-Zr | 30-40 | -        | -                      | 0.04±0.01                 | 2.0±0.2                 |
| Sn-Ba | 50-56 | -        | -                      | 0.0066±0.001              | 2.0±2                   |
| Os-Pb | 76-82 | -        | -                      | 0.0007±.0003              | 0.64±0.12               |

[1] Lodders, Palme, and Gail (2009), see also Asplund *et al.* (2009).

[*] Caffau *et al.* (2011).

[2] Reames (1995, 2014, 2017a).

[3] Reames, Cliver, and Kahler (2014a)

[**] Reames (2019a).

# References

Asplund, M., Grevesse, N., Sauval, A.J., Scott, P.: 2009, The chemical composition of the sun, *Ann. Rev. Astron. Astrophys*. **47**, 481 doi: 10.1146/annurev.astro.46.060407.145222

Breneman, H.H., Stone, E.C.: 1985, Solar coronal and photospheric abundances from solar energetic particle measurements, *Astrophys. J. Lett.* **299**, L57, doi: 10.1086/184580

Bučík, R., Innes, D.E., Chen, N.H., Mason, G.M., Gómez-Herrero, R., Wiedenbeck, M.E.: 2015, Long-lived energetic particle source regions on the Sun, *J. Phys. Conf. Ser.* **642**, 012002  doi: 10.1088/1742-6596/642/1/012002

Bučík,,R,. Innes, D.E., Mall, U., Korth, A., Mason, G.M., Gómez-Herrero, R.: 2014, Multi-spacecraft observations of recurrent $^3$He-rich solar energetic particles, *Astrophys. J.* **786**, 71 doi: 10.1088/0004-637X/786/1/71

Bučík, R., Innes, D.E., Mason, G.M., Wiedenbeck, M.E., Gómez-Herrero, R., Nitta, N.V.:2018, $^3$He-rich solar energetic particles in helical jets on the Sun, *Astrophys. J.* **852** 76 doi: 10.3847/1538-4357/aa9d8f

Caffau, E., Ludwig, H.-G., Steffen, M., Freytag, B., Bonofacio, P.: 2011, Solar chemical abundances determined with a CO5BOLD 3D model atmosphere, *Solar Phys*. **268**, 255. doi:10.1007/s11207-010-9541-4

Chen, N.H., Bučík R., Innes D.E., Mason G.M. 2015, Case studies of multi-day $^3$He-rich solar energetic particle periods, *Astron. Astrophys*. **580**, 16 doi: 10.1051/0004-6361/201525618

Cliver, E.W., Kahler, S.W., Reames, D.V.:2004, Coronal shocks and solar energetic proton events, *Astrophys. J.* **605** 902 doi: 10.1086/382651

Cohen, C.M.S., Mason, G.M., Mewaldt, R.A.: 2017, Characteristics of solar energetic ions as a function of longitude, *Astrophys. J.* **843**, 132 doi: 10.3847/1538-4357/aa7513

Desai, M.I., Giacalone, J.: 2016, Large gradual solar energetic particle events, *Living Reviews of Solar Physics*, doi: 10.1007/s41116-016-0002-5.

Desai, M.I., Mason, G.M., Dwyer, J.R., Mazur, J.E., Gold, R.E., Krimigis, S.M., Smith, C.W., Skoug, R.M.: 2003, Evidence for a suprathermal seed population of heavy ions accelerated by interplanetary shocks near 1 AU, *Astrophys. J.* **588**, 1149, doi: 10.1086/374310

Drake, J.F., Cassak, P.A., Shay, M.A., Swisdak, M., Quataert, E.:2009, A magnetic reconnection mechanism for ion acceleration and abundance enhancements in impulsive flares, *Astrophys. J. Lett.* **700**, L16 doi: 10.1088/0004-637X/700/1/L16

Gosling, J.T.: 1993, The solar flare myth, *J. Geophys. Res.* **98**, 18937 doi: 10.1029/93JA01896

Kahler, S.W., Reames, D.V., Sheeley, N.R.Jr.: 2001, Coronal mass ejections associated with impulsive solar energetic particle events, *Astrophys. J.,* **562**, 558  doi: 10.1086/323847

Kahler, S.W., Sheeley, N.R.Jr., Howard, R.A., Koomen, M.J., Michels, D.J., McGuire R.E., von Rosenvinge, T.T., Reames, D.V.: 1984, Associations between coronal mass ejections and solar energetic proton events, *J. Geophys. Res.* **89**, 9683, doi: 10.1029/JA089iA11p09683

Laming, J.M.: 2009, Non-WKB models of the first ionization potential effect: implications for solar coronal heating and the coronal helium and neon abundances, *Astrophys. J.* **695**, 954 doi: 10.1088/0004-637X/695/2/954

Laming, J.M.: 2015, The FIP and inverse FIP effects in solar and stellar coronae, *Living Reviews in Solar Physics*, **12**, 2 doi: 10.1007/lrsp-2015-2

Lee, M.A.: 1983, Coupled hydromagnetic wave excitation and ion acceleration at interplanetary traveling shocks*, J. Geophys. Res.,* **88**, 6109  doi: 10.1029/JA088iA08p06109

Lee, M.A.: 2005, Coupled hydromagnetic wave excitation and ion acceleration at an evolving coronal/interplanetary shock, *Astrophys. J. Suppl.,* **158**, 38 doi: 10.1086/428753

Lee, M.A., Mewaldt, R.A., Giacalone, J.: 2012, Shock acceleration of ions in the heliosphere, *Space Sci. Rev.,* **173**, 247, doi: 10.1007/s11214-012-9932-y

Lodders, K., Palme, H., Gail, H.-P.: 2009, Abundances of the elements in the solar system, In: Trümper, J.E. (ed.) *Landolt-Börnstein, New Series VI/4B*, Springer, Berlin. Chap. **4.4**, 560.

Mason, G. M.: 2007 $^3$He-rich solar energetic particle events, *Space Sci. Rev*. **130**, 231 doi: 10.1007/s11214-007-9156-8

Mason, G.M., Mazur, J.E., Dwyer, J.R.: 2002, A new heavy ion abundance enrichment pattern in $^3$He-rich solar particle events. *Astrophys. J. Lett*. 565, L51 doi: 10.1086/339135

Mason, G.M., Mazur, J.E., Dwyer, J.R., Jokipii, J.R., Gold, R.E., Krimigis, S.M.: 2004, Abundances of heavy and ultra-heavy ions in $^3$He-rich solar flares, *Astrophys. J.* **606**, 555 doi: 10.1086/382864

Mason, G.M., Ng, C.K., Klecker, B., Green, G.: 1989, Impulsive acceleration and scatter-free transport of about 1 MeV per nucleon ions in $^3$He-rich solar particle events, *Astrophys. J*. **339**, 529 doi: 10.1086/167315

Mewaldt, R.A., Cohen, C.M.S., Leske, R.A., Christian, E.R., Cummings, A.C., Stone, E.C., von Rosenvinge, T.T., Wiedenbeck, M.E.: 2002, Fractionation of solar energetic particles and solar wind according to first ionization potential, *Adv. Space Res.***30** 79 doi: 10.1016/S0273-1177(02)00263-6

Meyer, J.-P.: 1985, The baseline composition of solar energetic particles, *Astrophys. J. Suppl*. **57**, 151, doi: 10.1086/191000

Ng, C.K., Reames, D.V.: 2008, Shock acceleration of solar energetic protons: the first 10 minutes, *Astrophys. J. Lett*. **686**, L123 doi: 10.1086/592996

Ng, C.K., Reames, D.V., Tylka, A.J.: 1999, Effect of proton-amplified waves on the evolution of solar energetic particle composition in gradual events, *Geophys. Res. Lett*. **26**, 2145 doi: 10.1029/1999GL900459

Ng, C.K., Reames, D.V., Tylka, A.J.: 2001, Abundances, spectra, and anisotropies in the 1998 Sep 30 and 2000 Apr 4 large SEP events, *Proc.27$^{th}$ Int. Cosmic-Ray Conf.* (Hamburg), **8**, 3140

Ng, C.K., Reames, D.V., Tylka, A.J.: 2003, Modeling shock-accelerated solar energetic particles coupled to interplanetary Alfvén waves, *Astrophys. J*. **591**, 461, doi: 10.1086/375293

Ng, C.K., Reames, D.V., Tylka, A.J.: 2012, Solar energetic particles: Shock acceleration and transport through self-amplified waves, **CP-1436** AIP, Melville 212, doi: 10.1063/1.4723610

Parker, E.N.: 1963 *Interplanetary Dynamical Processes,* Interscience, New York.

Reames, D.V.: 1988, Bimodal abundances in the energetic particles of solar and interplanetary origin, *Astrophys. J. Lett*. **330**, L71 doi: 10.1086/185207

Reames, D.V.: 1995, Coronal Abundances determined from energetic particles, *Adv. Space Res.* **15** (7), 41.

Reames, D.V.: 1999, Particle acceleration at the sun and in the heliosphere, *Space Sci. Rev*. **90**, 413 doi: 10.1023/A:1005105831781

Reames, D.V.: 2000, Abundances of trans-iron elements in solar energetic particle events, *Astrophys. J. Lett*. **540**, L111 doi: 10.1086/312886

Reames, D.V.:2009, Solar release times of energetic particles in ground-level events, *Astrophys. J.* **693**, 812 doi: 10.1088/0004-637X/693/1/812

Reames, D.V.: 2013, The two sources of solar energetic particles, *Space Sci. Rev.* **175**, 53, doi: 10.1007/s11214-013-9958-9

Reames, D.V.:2014, Element abundances in solar energetic particles and the solar corona, *Solar Phys*. **289**, 977 doi: 10.1007/s11207-013-0350-4

Reames, D.V.: 2015, What are the sources of solar energetic particles? Element abundances and source plasma temperatures, *Space Sci. Rev.,* **194**: 303, doi: 10.1007/s11214-015-0210-7.

Reames, D.V.: 2016a, Temperature of the source plasma in gradual solar energetic particle events, *Solar Phys*. **291,** 911, doi: 10.1007/s11207-016-0854-9

Reames, D.V.:2016b, The origin of element abundance variations in solar energetic particles, *Solar Phys*. **291,** 2099, doi: 10.1007/s11207-016-0942-x,

Reames D.V.: 2017a, *Solar Energetic Particles*, Lecture Notes in Physics **932**, Springer, Berlin, ISBN 978-3-319-50870-2, doi: 10.1007/978-3-319-50871-9.

Reames, D.V.: 2017c, The abundance of helium in the source plasma of solar energetic particles, *Solar Phys*. **292,** 156 doi: 10.1007/s11207-017-1173-5 (arXiv: 1708.05034)

Reames, D.V., 2018a The "FIP effect" and the origins of solar energetic particles and of the solar wind, Solar Phys. **293,** 47 doi: 10.1007/s11207-018-1267-8 (arXiv 1801.05840 )

Reames, D.V.: 2018b, Abundances, ionization states, temperatures, and FIP in solar energetic particles, *Space Sci. Rev*. **214,** 61, doi: 10.1007/s11214-018-0495-4 (arXiv: 1709.00741)

Reames, D.V.: 2019a, Helium suppression in impulsive solar energetic-particle events, *Solar Phys.* in press, doi: 10.1007/s11207-019-1422-x (arXiv: 1812.01635)

Reames, D.V.: 2019b, Hydrogen and the abundances of elements in gradual solar energetic-particle events, *Solar Phys.* submitted (arXiv 1902.03208 )

Reames, D.V., Ng, C.K.: 2004, Heavy-element abundances in solar energetic particle events, *Astrophys. J.* **610**, 510 doi: 10.1086/421518

Reames, D.V., Ng, C.K.: 2010, Streaming-limited intensities of solar energetic particles on the intensity plateau, *Astrophys. J.* **723** 1286 doi: 10.1088/0004-637X/723/2/1286

Reames, D.V., Stone, R.G.: 1986, The identification of solar $^3$He-rich events and the study of particle acceleration at the sun, *Astrophys. J.,* **308**, 902 doi:  10.1086/164560

Reames, D.V., Barbier, L.M., Ng, C.K.,:1996, The spatial distribution of particles accelerated by coronal mass ejection-driven shocks, *Astrophys. J.* **466** 473 doi: 10.1086/177525

Reames, D.V., Barbier, L.M., von Rosenvinge, T.T., Mason, G.M., Mazur, J.E., Dwyer, J.R.: 1997, Energy spectra of ions accelerated in impulsive and gradual solar events, *Astrophys. J.* **483**, 515 doi: 10.1086/304229

Reames, D.V., Cliver, E.W., Kahler, S.W.: 2014a, Abundance enhancements in impulsive solar energetic-particle events with associated coronal mass ejections, *Solar Phys*. **289**, 3817 doi: 10.1007/s11207-014-0547-1

Reames, D.V., Cliver, E.W., Kahler, S.W.: 2014b Variations in abundance enhancements in impulsive solar energetic-particle events and related CMEs and flares, *Solar Phys.* **289**, 4675 doi: 10.1007/s11207-014-0589-4

Reames, D.V., Cliver, E.W., Kahler, S.W.: 2015, Temperature of the source plasma for impulsive solar energetic particles, *Solar Phys*. **290**, 1761 doi: 10.1007/s11207-015-0711-2

Reames, D.V., Kahler, S.W., Ng, C.K.: 1997, Spatial and temporal invariance in the spectra of energetic particles in gradual solar events, *Astrophys. J.* **491**, 414 doi: 10.1086/304939

Reames, D.V., Meyer, J.P., von Rosenvinge, T.T.: 1994, Energetic-particle abundances in impulsive solar flare events, *Astrophys. J. Suppl.* **90**, 649 doi: 10.1086/191887

Reames, D.V., Ng, C.K., Berdichevsky, D., 2001, Angular Distributions of Solar Energetic Particles, *Astrophys. J.* **550** 1064 doi: 10.1086/319810

Reames, D.V., Ng, C.K., Tylka, A.J.: 2000, Initial time dependence of abundances in solar particle events, *Astrophys. J. Lett*. **531,** L83  doi: 10.1086/312517

Reames, D.V., von Rosenvinge, T.T., Lin, R.P.: 1985, Solar $^3$He-rich events and nonrelativistic electron events - A new association, *Astrophys. J*. **292**, 716 doi: 10.1086/163203

Roth, I., Temerin, M.: 1997, Enrichment of $^3$He and Heavy Ions in Impulsive Solar Flares, *Astrophys. J.* **477**, 940 doi: 10.1086/303731

Rouillard, A., Sheeley, N.R. Jr., Tylka, A., Vourlidas, A., Ng, C.K., Rakowski, C., Cohen, C.M.S., Mewaldt, R.A., Mason, G.M., Reames, D., *et al.*: 2012, The longitudinal properties of a solar energetic particle event investigated using modern solar imaging, *Astrophys. J*. **752**, 44 doi: 10.1088/0004-637X/752/1/44

Schmelz , J. T., Reames, D. V., von Steiger, R., Basu, S.: 2012, Composition of the solar corona, solar wind, and solar energetic particles, *Astrophys. J.* **755** 33 doi: 10.1088/0004-637X/755/1/33

Serlemitsos, A.T., Balasubrahmanyan, V.K.: 1975, Solar particle events with anomalously large relative abundance of $^3$He, *Astrophys. J.* **198**, 195 doi: 10.1086/153592

Temerin, M., Roth, I.: 1992, The production of $^3$He and heavy ion enrichment in $^3$He-rich flares by electromagnetic hydrogen cyclotron waves, *Astrophys. J. Lett*. **391**, L105 doi: 10.1086/186408

Tylka, A.J., Boberg, P.R., Cohen, C.M.S., Dietrich, W.F., Maclennan, C.G. Mason, G.M., Ng, C.K. Reames, D.V.: 2002, Flare- and shock-accelerated energetic particles in the solar events of 2001 April 14 and 15, *Astrophys. J.* **581** 119 doi: 10.1086/346033

Tylka, A.J., Cohen, C.M.S., Dietrich, W.F., Lee, M.A., Maclennan, C.G., Mewaldt, R.A., Ng, C.K., Reames, D.V.: 2005, Shock geometry, seed populations, and the origin of variable elemental composition at high energies in large gradual solar particle events, *Astrophys. J.* **625**, 474 doi: 10.1086/429384

Tylka, A.J., Lee, M.A.: 2006, Spectral and compositional characteristics of gradual and impulsive solar energetic particle events, *Astrophys. J.* **646**, 1319 doi: 10.1086/505106

Webber, W.R.: 1975, Solar and galactic cosmic ray abundances - A comparison and some comments. *Proc. 14$^{th}$ Int. Cosmic Ray Conf, Munich*, **5** 1597

von Rosenvinge, T.T., Barbier, L.M., Karsch, J., Liberman, R., Madden, M.P., Nolan, T., Reames, D.V., Ryan, L., Singh, S., Trexel, H.: 1995, The Energetic Particles: Acceleration, Composition, and Transport (EPACT) investigation on the *Wind* spacecraft, *Space Sci. Rev.* **71**  152 doi: 10.1007/BF00751329

Zank, G.P., Rice, W.K.M., Wu, C.C.: 2000, Particle acceleration and coronal mass ejection driven shocks: A theoretical model, *J. Geophys. Res.,* 105, 25079 doi: 10.1029/1999JA0004551